\documentclass[prb,preprint,aps,amssymb,epsfig]{revtex4}
\draft
\usepackage{graphicx}
\usepackage{subfigure}
\usepackage{amsfonts}
\usepackage{amssymb}
\usepackage{amsmath}
\begin{document}
\title{
Hydrogen supersymmetry: A new method in perturbation theory
}
\author{E. A.~Muljarov}
\address{General Physics Institute, Russian Academy of Sciences,
Vavilova 38, Moscow 119991, Russia}
\begin{abstract}
The {\it O(4)} supersymmetry of the hydrogen atom is utilized to
construct a complete basis using only the bound state wave
functions. For a large class of perturbations, an expansion of the
electron (exciton) wave function into such a complete set reduces
the perturbed Schr\"odinger equation to a standard eigenvalue
problem with an equidistant unperturbed spectrum. A high-order
polarizability of the ground state of the hydrogen atom in a static
electric field is calculated via Rayleigh-Schr\"odinger
perturbation theory for illustration of the new method.
\end{abstract}
 \maketitle
\section{Introduction}
\label{sec-intro}

The task of the perturbation theory is to calculate, at least
approximately, the energy spectrum and the wave functions of the
given problem using those of the already solved primary problem. As
a rule, the spectrum of the primary problem consists of discrete
and continuum states. Both of them should be taken into account in
the expansion of the perturbed wave function. However, in practice
the inclusion of the continuum is often an extremely complicated
task. This is the case, for instance, of the hydrogen problem,
where the bound state wave functions do not constitute a complete
set even approximately: Some of the scattering states are
energetically close to the bound ones, since the levels thickening
point of the latter meets the continuum onset.

In the present paper we show explicitly that a
specific inhomogeneous coordinate scaling applied to the hydrogen
bound-state wave functions converts them into a complete set. Early
in 1935 it was shown by V.A.~Fock that due to the hidden symmetry
of the Coulomb potential (which leads to the additional spectrum
degeneracy) the three-dimensional (3D) hydrogen problem in the
momentum space can be one-to-one mapped onto a sphere or onto a
two-sheeted hyperboloid in 4D space, for bound or scattering
states, respectively.\cite{Fock35} The full space of the hydrogen
eigenfunctions is thus divided into two subspaces, the
eigenfunctions of each subspace being an irreducible representation
of the corresponding symmetry group. In particular, the bound-state
wave functions are transformed into the hyperspherical harmonics, the
irreducible representation of the rotational symmetry group {\it
O(4)} in the 4D energy-momentum space.\cite{Band66}

An expansion into the complete basis of hyperspherical harmonics
(also referred to as Sturmian functions -- in the coordinate
representation) has been
already used in nuclear physics when treating a three-body problem
(see, e.g., Refs~\onlinecite{Rote62,Shak75}) and recently in
anisotropic exciton problem,\cite{Mulj99} when the perturbation
theory was constructed on a sphere in 4D space. In the present work
we formulate the perturbation theory directly in the coordinate
space, basing on the complete set of functions orthogonal with the
weight $1/r$. The important feature of this basis is that only {\em
normalizable} (square integrable) wave functions, which belong to the
bound states of the perturbed problem, can be expanded. Due to the
energy-dependent transformation of the standard Coulomb wave
functions which makes the basis complete, the Schr\"odinger
equation is reduced to the eigenvalue problem with equidistant
unperturbed energy spectrum. Finally, the Brillouin-Wigner
perturbation theory converges for any perturbation which only
renormalizes the Coulomb spectrum leaving a set of bound states
therein. The classical example of such a perturbation is the
anisotropic exciton Hamiltonian. In spite of its apparent
simplicity and enduring interest, the exact approach to the
anisotropic exciton problem based on the present method was
developed only recently.\cite{Mulj99} In other approaches the
expansion of the wave function into an incomplete basis (see, e.g.
Ref.~\onlinecite{Faul69}) led to substantial deviation in numerical
results compared to the exact method, at least for rather strong
anisotropy.\cite{Mulj99} Nevertheless, in this work we do not dwell
into details of concrete physical examples but only formulate the
general recipe how to treat such kind of problems.

As soon as the perturbation leads to drastic changes of the
spectrum, such as decay of formerly bound states, Brillouin-Wigner
perturbation theory is not convergent any more and the solution of
a truncated eigenvalue problem becomes meaningless. However, finite
orders of Rayleigh-Schr\"odinger perturbation theory still may have
some physical meaning. This is the case, e.g., of the hydrogen atom
(or Wannier exciton) in a static electric field which makes the
bound states of an electron (exciton) metastable, showing up in the
time decay of the wave functions. At the same time, the
polarizability of an atom or an exciton is described by the second
perturbation order.\cite{Land76} To illustrate how the method works
in such situations we calculate a high-order polarizability of the
ground state in a static electric field.


\section{Construction of the complete basis}
\label{sec-basis}

The Schr\"odinger equation of the hydrogen atom has the form
\begin{equation}
\left(-\frac{\hbar^2}{2\mu}\nabla^2 -\frac{e^2}{r} - E\right)
\psi({\bf r})=0,
\label{Schr-Hydrogen}
\end{equation}
where $E$, ${\bf r}$, $\mu$, and $e$ are the electron energy,
coordinate, mass, and the elementary charge ($e>0$), respectively.
Following Fock,\cite{Fock35} let us introduce for the bound states
$E<0$ the energy-dependent parameter
\begin{equation}
a=\frac{1}{2}\sqrt{\frac{\hbar^2/2\mu}{-E}}
\label{a-def}
\end{equation}
and make the following coordinate transformation
$
\mbox{\boldmath$\rho$}={\bf r}/a$, where
$\mbox{\boldmath$\rho$}=\rho\cdot(\sin\theta\cos\varphi,
\sin\theta\cos\varphi, \cos\theta)$.
Then, introducing a new wave function as
$
\phi(\mbox{\boldmath$\rho$})\propto \sqrt{\rho }\,\psi({\bf r}),
$
we result in the following differential equation
\begin{equation}
\left\{
-\frac{\partial}{\partial \rho} \rho \frac{\partial}{\partial \rho}
+\frac{1}{\rho}\left[\frac{1}{4}-\hat{\Lambda}(\theta,\varphi)\right]
+\frac{\rho}{4}
-\lambda\right\}\phi(\mbox{\boldmath$\rho$})=0,
\label{dif-equ}
\end{equation}
where
$\hat{\Lambda}(\theta,\varphi)$
is the angular part of the Laplace operator,
\begin{equation}
\lambda=2\frac{a}{a_B}=\sqrt{\frac{\hbar^2/2\mu a_B^2}{-E}}
\label{lambda-def}
\end{equation}
is a new dimensionless eigenvalue which originates from the Coulomb
term in Eq.~(\ref{Schr-Hydrogen}), $a_B=\hbar^2/\mu e^2$ is the Bohr
radius.

Equation~(\ref{dif-equ}) has the following
eigensolutions\cite{Grad80}
\begin{equation}
\phi_{nlm}(\mbox{\boldmath$\rho$})=
y_{nl}(\rho)\,{\cal Y}_{lm}(\theta,\varphi),
\label{phi-fun}
\end{equation}
where ${\cal Y}_{lm}$ are the standard spherical harmonics,
$\{y_{nl}\}$ is the orthogonal set of functions
\begin{equation}
y_{nl}(\rho)=\sqrt{\frac{(n-l-1)!}{(n+l)!}}
\rho^{l+1/2} e^{-\frac{\rho}{2}} L_{n-l-1}^{2l+1} (\rho)
\label{y-fun}
\end{equation}
normalized as
\begin{equation}
\int\limits_0^\infty d\rho \,y_{n'l}(\rho) y_{nl}(\rho)
= \delta _{nn'},
\label{norm}
\end{equation}
$L_p^q$ are the generalized Laguerre polynomials (for definition
see, e.g., Ref.~\onlinecite{Grad80}), ($n,l,m)$ are the standard
hydrogen quantum numbers. The eigenvalues of Eq.~(\ref{dif-equ})
have the following form\cite{fn1}
\begin{equation}
\lambda^{(0)}_{nlm}=n,\ \ \ n=1,\,2,\,\dots,
\label{lambda0}
\end{equation}
leaving the same degeneracy as of the usual hydrogen spectrum.
Accordingly, the scaling parameter $a$ depends on the principle
quantum number as $a=n\,a_B/2$
that finally results in the standard Coulomb wave functions,
normalized solutions of Eq.~(\ref{Schr-Hydrogen}),
\begin{equation}
\psi^C_{nlm}({\bf r})=\sqrt{\frac{n a_B}{2r}}
\phi_{nlm}\left(\frac{2{\bf r}}{n a_B}\right).
\end{equation}

Being the full solution of the radial wave equation following from
Eq.~(\ref{dif-equ}), the set of functions $\{y_{nl}(\rho)\}$,
Eq.~(\ref{y-fun}), also referred to as Sturmian basis,\cite{Rote62}
constitutes a complete basis in the one-dimensional $\rho$-space.
At the same time, the spherical harmonics ${\cal Y}_{lm}$ are the
complete set of functions on a sphere. Consequently, the product of
these two, the functions $\phi_{nlm}(\mbox{\boldmath$\rho$})$,
Eq.~(\ref{phi-fun}), although they are not orthogonal to each
other, make up a complete set in the three-dimensional
\mbox{\boldmath$\rho$}-space. If we now treat $a$ as a {\em
constant parameter}, one and the same for all $\phi_{nlm}$, we will
result in the complete set of functions in the electron coordinate
space,
\begin{equation}
\chi_{nlm}({\bf r};a)= \frac{1}{\sqrt{ r a}} \,y_{nl}\left({r}/{a}\right)
{\cal Y}_{lm}(\theta,\varphi)= \sqrt{\frac{2}{n a a_B}} \,\psi^C_{nlm}
\left(\frac{n a_B}{2a}{\bf r}\right),
\label{chi-def}
\end{equation}
normalized as
\begin{equation}
\int \frac{d{\bf r}}{r}\chi^\ast_\nu({\bf r};a) \chi_{\nu'}({\bf
r};a)=\delta_{\nu\nu'}, \ \ \ \nu=(n,l,m).
\label{orthog}
\end{equation}

Looking at the basic functions Eq.~(\ref{chi-def}) one can notice
that, contrary to the Coulomb wave functions, all $\chi_\nu$ have
one and the same exponential decay factor $e^{-r/2a}$,
the remainder being a set of orthogonal
polynomials of $r$. Any square-integrable function, which decays at ${\bf
r}\to\infty$, can be expanded into the set $\{\chi_\nu({\bf
r};a)\}$.

\section{Formulation of the Brillouin-Wigner perturbation theory}
\label{sec-ptheory}

Let us now consider the Schr\"odinger equation with a perturbation
$\hat{U}$ which can in principle contain differential and/or
integration operators,
\begin{equation}
\left[-\frac{\hbar^2}{2\mu}\nabla^2 -\frac{e^2}{r} + \hat{U}({\bf r})-
E\right]
\Psi({\bf r})=0.
\label{Schr-pert}
\end{equation}
Expand the wave function as
\begin{equation}
\Psi({\bf r})=\sum_\nu C_\nu \chi_\nu({\bf r};a).
\label{expansion}
\end{equation}
Let's define again the basic parameter $a$ as in Eq.~(\ref{a-def})
with $E$ being now the energy of the perturbed state. Then we have
\begin{equation}
\left(-\frac{\hbar^2}{2\mu}\nabla^2 -\frac{e^2}{r} - E\right)\chi_\nu({\bf
r};a)=
\frac{\hbar^2}{2\mu a}(n-\lambda)\frac{1}{r}\,\chi_{nlm}({\bf r};a),
\label{n-lam}
\end{equation}
where
$\lambda$ is defined by Eq.~(\ref{lambda-def}).

Plugging Eq.~(\ref{expansion}) into Eq.~(\ref{Schr-pert}), using
Eq.~(\ref{n-lam}) and the orthogonality condition
Eq.~(\ref{orthog}), after convolution with $\chi_{\nu'}$ we result
in the following matrix equation
\begin{equation}
\sum_{\nu'} \Bigl[n\delta_{\nu\nu'}+ V_{\nu\nu'}\Bigr]C_{\nu'}=\lambda C_\nu,
\label{ev-problem}
\end{equation}
with the unperturbed spectrum defined by Eq.~(\ref{lambda0}) and the
perturbation matrix
\begin{equation}
V_{\nu\nu'}=\frac{2\mu a}{\hbar^2}\int  \chi^\ast_\nu({\bf r};a)
\hat{U}({\bf r})\chi_{\nu'}({\bf r};a)d{\bf r}.
\label{me-first}
\end{equation}
At first glance, Eq.~(\ref{ev-problem}) looks like a generalized
eigenvalue problem, since the definition of $V_{\nu\nu'}$ contains
the energy-dependent parameter $a$, and it seems that it cannot be
solved by means of the direct diagonalization of the effective
Hamiltonian. However, it is easily seen that any perturbation
$\hat{U}$ can be considered as an operator which is governed by a
set of physical parameters $\alpha_1,\alpha_2,\dots$, such as
external field strength, anisotropy degree, coupling constant, and
so on: $\hat{U}({\bf r};\alpha_1,\alpha_2,\dots)$. After the
coordinate transformation ${\bf r}=\mbox{\boldmath$\rho$}a$ it can
be rewritten as
\begin{equation}
\frac{2\mu a^2}{\hbar^2}\hat{U}({\bf r};\alpha_1,\alpha_2,\dots)=
\hat{V}(\mbox{\boldmath$\rho$};\beta_1,\beta_2,\dots),
\label{U-V}
\end{equation}
where the operator $\hat{V}$ does not depend explicitly on $a$ but
is a function of new parameters
$ \beta_i=\beta_i(a;\alpha_1,\alpha_2,\dots)$,
which, in turn, depend on $a$. The matrix elements
Eq.~(\ref{me-first}) now take the form [see the definition of
$\phi_\nu$, Eq.~(\ref{phi-fun})]
\begin{equation}
V_{\nu\nu'}(\beta_1,\beta_2,\dots)=\int
\frac{d\mbox{\boldmath$\rho$}}{\rho}
\phi^\ast_\nu(\mbox{\boldmath$\rho$})
\hat{V}(\mbox{\boldmath$\rho$};\beta_1,\beta_2,\dots)
\phi_{\nu'}(\mbox{\boldmath$\rho$}).
\end{equation}
Treating $\beta_i$ (instead of $\alpha_i$) as fixed parameters of
the Brillouin-Wigner perturbation theory, we finally get rid of the
$a$-dependence in Eq.~(\ref{ev-problem}). Thus,
Eq.~(\ref{ev-problem}) becomes the ordinary eigenvalue problem. The
dependence of the spectrum on the physical parameters $\alpha_i$
can be restored by means of inverse functions $
\alpha_i=\alpha_i(a;\beta_1,\beta_2,\dots)$.
Obviously, such back transformation exists only if there is
one-to-one relation between $\alpha$'s and $\beta$'s. In other
words, $\alpha_i$ should be a single-valued function. This imposes
some additional restrictions on the perturbation $\hat{U}$.
Moreover, the requirement to have bound states in the perturbed
spectrum implies the asymptotic behavior $\hat{U}({\bf r})\to 0$
at $r\to\infty$. In the next section we however consider an example
of potential which does not satisfy the latter condition.

\section{Example: Hydrogen atom in a static electric field}
\label{sec-polariz}
The interaction of an electron with a static electric field
$\mbox{\boldmath${\cal E}$}=(0,0,{\cal E})$ has the form
\begin{equation}
\hat{U}({\bf r})=-z e{\cal E}.
\label{U-ef}
\end{equation}
According to Eq.~(\ref{U-V}), the corresponding dimensionless
operator is
\begin{equation}
\hat{V}(\mbox{\boldmath$\rho$};\beta)=-\beta \rho^2 \cos\theta
\label{V-pert}
\end{equation}
where
\begin{equation}
\beta=\frac{2\mu a^3}{\hbar^2} e{\cal E}.
\label{beta-E}
\end{equation}
The matrix elements $V_{\nu\nu'}$ are calculated in
Appendix~\ref{sec-app}. They have simple analytical form and
satisfy rigorous selection rules. Namely, only the elements with
\begin{equation}
m'=m,\ \ \ l'=l\pm1,\ \ \ n'=n,\,n\pm1,\,n\pm2
\end{equation}
are nonvanishing.

As it is already mentioned, the Brillouin-Wigner perturbation
theory fails when the perturbation destroys the bound states
spectrum. In case of the perturbation Eq.~(\ref{U-ef}), the
electric field potential provides a non-zero probability of an
electron to tunnel trough the local Coulomb barrier. The states
become quasi-bound, having nonvanishing imaginary part of the
energy. On the other hand, the Rayleigh-Schr\"odinger perturbation
series is Borel summable, having zero radius of
convergence.\cite{Avro77} Although the required accuracy can be
achieved by sufficient number of perturbation orders taken into
account, the higher the accuracy of calculations one wants to
approach the smaller should be the region of the field magnitude.

Let's calculate corrections to the electron ground state energy,
accounting for only the first few orders of the perturbation
theory. As the electron ground state is non-degenerate, the $n$-th
order can be calculated by means of the following recurrent
formula\cite{Elyu76}
\begin{eqnarray}
\lambda&=&\sum_s \lambda^{(s)},
\ \ \
\lambda^{(s)}=\sum_\nu  V_{0\nu} C^{(s-1)}_{\nu}
\nonumber\\
C^{(0)}_{\nu}&=&\delta_{\nu0},
\ \ \
C^{(s)}_{0}=0 \ \ \ (s>0),
\nonumber\\
C^{(s)}_{\nu}&=&\frac{1}{\lambda_\nu^{(0)}-\lambda^{(0)}}
\left[ -\sum_{\nu'} V_{\nu\nu'}C^{(s-1)}_{\nu'}
+\sum_{r=1}^{s-1} \lambda^{(r)}C^{(s-r)}_{\nu}\right] \ \ \
(s>0,\ \nu>0),
\end{eqnarray}
where the upper script shows the perturbation order in the
calculation of the ground state eigenvalue and eigenvector. For
simplicity, in eigenvectors and matrix elements $\nu=0$ denotes the
unperturbed ground state; it is omitted in eigenvalues
$\lambda^{(s)}\equiv\lambda_0^{(s)}$.

Due to the fact that $\hat{V}$ couples any unperturbed state with
only finite number of states, the eigenvalues and eigenvectors of
each finite order can be easily calculated analytically. For
example, up to the eighth order, the ground state eigenvalue has
the form
\begin{eqnarray}
\lambda=1-36\,\beta^2-20\,052\,\beta^4-292\,973\,512\,\beta^6
-73\,522\,897\,716\,\beta^8-\dots
\label{lambda-eight}
\end{eqnarray}
To return back to the energy and electric field we use
Eqs.~(\ref{a-def}), (\ref{lambda-def}), and (\ref{beta-E}) that
yields an indirect dependence $E({\cal E})$. The numerical plots
of the ground state energy measured in Rydberg versus the electric
field strength measured in ${\cal E}_0=a_B^2/e$ are shown in
Fig.~1, for different perturbation orders $s$ taken into account.
Note that even the curve $s=2$ is not purely parabolic but contains
all orders of ${\cal E}$, since for any finite order $E$ is a
transcendental function of ${\cal E}$. Nevertheless, this function
can be expressed as a finite series of ${\cal E}$, all higher
orders being cut off. Then we result in the known
series\cite{Jent01}
\begin{equation}
\frac{E}{{\rm Ry}}=-1
-\frac{9}{2} \left(\frac{{\cal E}}{{\cal E}_0}\right)^2
-\frac{3\,555}{32} \left(\frac{{\cal E}}{{\cal E}_0}\right)^4
-\frac{2\,512\,779}{256}\left(\frac{{\cal E}}{{\cal E}_0}\right)^6
-\frac{13\,012\,777\,803}{8\,192}\left(\frac{{\cal E}}{{\cal E}_0}\right)^8
-\dots
\label{E-E}
\end{equation}
At first glance, this expression gives worse approximation than
Eq.~(\ref{lambda-eight}), since the energies are shifted upwards, see
dashed curves in the inset (Fig.~1). On the other hand, neither of
these two series is convergent at higher orders and it makes sense
only to speak about conditional convergence. As seen from Fig.~1,
at ${\cal E}>0.05{\cal E}_0$, the series Eq.~(\ref{lambda-eight})
\begin{figure}[t]
\vskip-1.3cm
\hskip-6cm
\includegraphics[angle=270,width=0.7\linewidth]{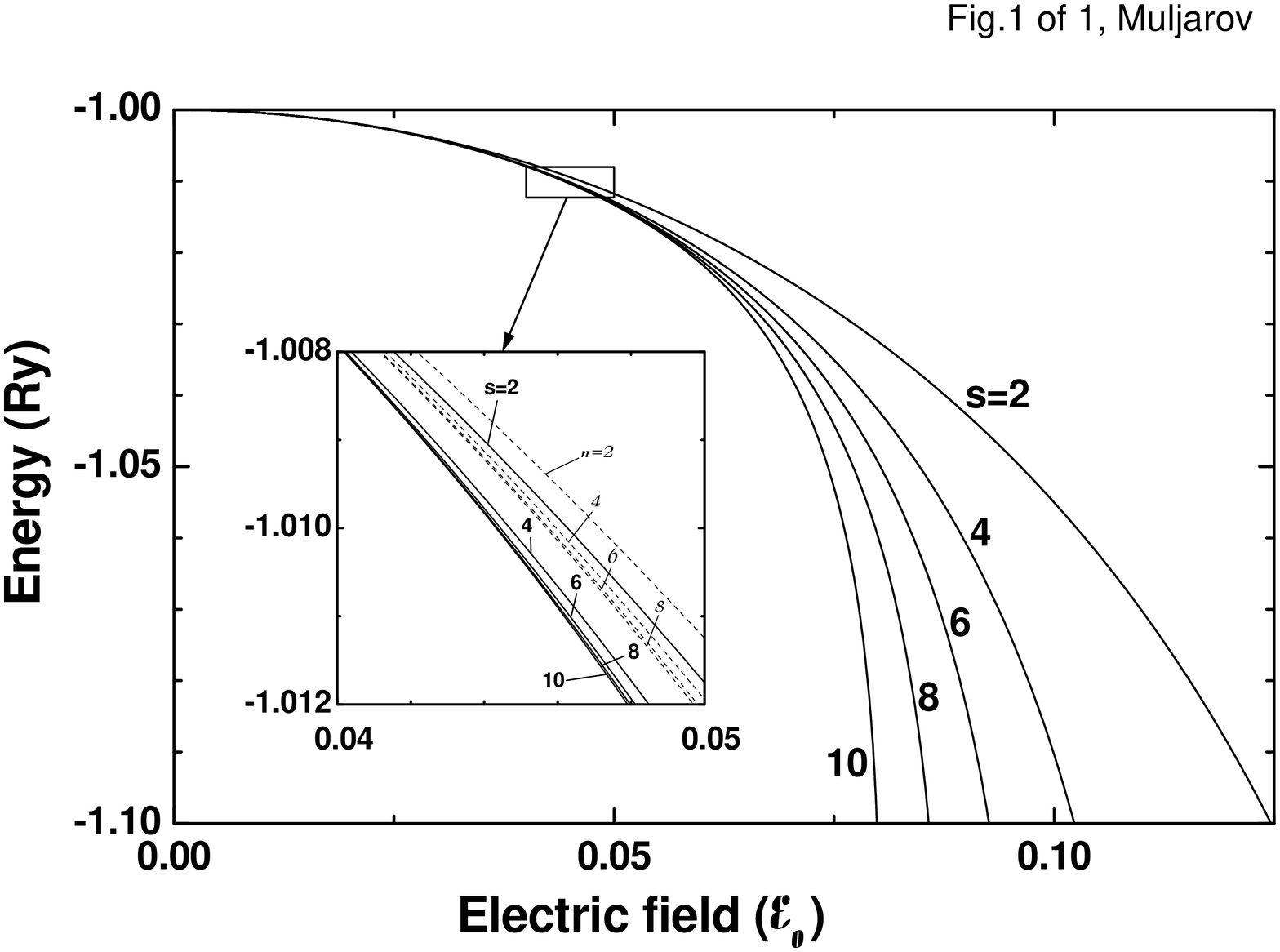}
\vskip-4cm
\caption{The ground state electron energy (measured in Rydberg) as a function
of the electric field (measured in ${\cal qE}_0=a_B^2/e$) calculated
via the perturbation theory up to $s=2$, 4, 6, 8, and 10th order
(full curves). The same calculated as a power series of ${\cal E}$
up to $n=2$, 4, 6, and 8th order (dashed curves). }
\end{figure}
is not convergent even for the lowest orders $s=2$ to 10: The
discrepancy in the energy calculated in two neighboring
perturbation orders does not become notably smaller with every next
order. In reality, the difference between Eq.~(\ref{lambda-eight})
and Eq.~(\ref{E-E}) simply estimates the accuracy of the
perturbative approach, which, in turn, is of the order of the
imaginary part of the complex eigenenergy.\cite{Jent01} At small
field strengths such decay rate of the bound electron (exciton)
state is rather small and one can get from the perturbation theory
a reasonable correction (of the order of several percent) to the
bound state energy.

\section{Discussion}
\label{sec-disc}

The new perturbation method formulated in the present
paper has a number of advantages compared to the standard
perturbation scheme. First of all, it allows to exclude from
consideration the continuum states of the unperturbed hydrogen-like
problem and, for a large class of perturbations, to reduce the
Schr\"odinger equation to the ordinary eigenvalue problem with
equidistant (instead of $-{\rm Ry}/n^2$) unperturbed 
spectrum. In particular, the discrete representation of the Coulomb
Green's function has the form
\begin{equation}
G_E({\bf r},{\bf r}')=\frac{2\mu a}{\hbar^2}\sum_{nlm}
\frac{1}{n-\lambda}
\chi^\ast_{nlm}({\bf r}';a) \chi_{nlm}({\bf r};a), \ \ \ E<0,
\label{Green-fun}
\end{equation}
where $\lambda(E)$ and $a(E)$ are given by Eqs.~(\ref{lambda-def})
and (\ref{a-def}), respectively. In contrast to the results of
Ref.~\onlinecite{Kony00} where the Laguerre-Sturmian expansion
leads to a tridiagonal form of the Green's operator, the Green's
function Eq.~(\ref{Green-fun}) has the standard diagonal form with
the discrete eigenvalues $\lambda^{(0)}=n$ snading in the
denominator.

Second, the use of the energy-dependent parameter $a$ in the
expansion of any square integrable function leads to one and
the same exponential decay of the basic functions, the correct
asymptotic of the expanded function. At
the same time, in the standard perturbative methods each basic
function has its own asymptotic behavior at $r\to\infty$ that
basically results in a slower convergence compared to the present
method.

Third, the matrix elements of the perturbation operator, after it is
transformed to the dimensionless form Eq.~(\ref{U-V}), are very
simple and in many cases can be calculated analytically.

Forth, the basic functions support the spherical symmetry that can
be very useful when a perturbation also possesses some symmetry.
For instance, the perturbative approach to the Stark problem,
Sec.~\ref{sec-polariz}, is very similar to the solution of the
problem in the parabolic coordinates.\cite{Land76} However,
introducing of the magnetic field into the Hamiltonian, makes the
`parabolic approach' extremely hard, since the basic functions are
not the eigenfunction of the angular momentum operators.
Contrarily, in our method, the matrix elements of the electron
potential in both electric and magnetic fields have simple
analytical form and selection rules.

Fifth, the complete set of functions Eq.~(\ref{chi-def}) can be
used for expansion of the wave function even if the Hamiltonian does
not contain Coulomb potential. One way to deal with such problems
is to add and subtract the Coulomb potential from the Hamiltonian
and and then follow the present perturbation method. Another way is to
expand the wave function of interest into the basis
Eq.~(\ref{chi-def}) with constant (energy-independent)
parameter $a$ and then
diagonalize the Hamiltonian.\cite{Rote62} In the latter case it is
recommended to treat $a$ as an adjustable parameter.

\acknowledgments

The author is grateful to Roland Zimmermann for helpful advices and
frequent discussions. The present work has been inspired by him
from the very beginning and it is a special pleasure for the author to
contribute to the Festschrift in his honor.
Also thanks to S.G.~Tikhodeev for critical reading of the
manuscript. The work is supported by the Russian Foundation for
Basic Research and Russian Ministry of Science (programs
``Nanostructures'' and ``Information Systems'').

\appendix
\section{Calculation of the matrix elements}
\label{sec-app}

The matrix elements of the perturbation $\hat{V}$,
Eq.~(\ref{V-pert}), have the form
\begin{equation}
V_{\nu \nu'} =-\beta \int \frac{d\mbox{\boldmath$\rho$}}{\rho}
\phi^\ast_\nu(\mbox{\boldmath$\rho$})\, \rho^2 \cos\theta\,
\phi_{\nu'}(\mbox{\boldmath$\rho$})=-\beta \delta_{mm'} J^m_{ll'}
I^{ll'}_{nn'},
\end{equation}
where ($m=0$)
\begin{equation}
J^0_{ll'}=\int d \Omega\, {\cal Y}_{l0}(\theta) \cos\theta\,
{\cal Y}_{l'0}(\theta)
=\frac{\sqrt{(2l+1)(2l'+1)}}{2} \int_{-1}^1 dt\, t P_l(t) P_{l'}(t),
\end{equation}
\begin{equation}
I_{nn'}^{ll'}\!=\!
\int\limits_0^\infty\!\! d\rho \,y_{nl}(\rho) \rho^2 y_{n'l'}(\rho)=
 \sqrt{\frac{(n-l-1)!(n'-l'-1)!}{(n+l)!(n'+l')!}} \int\limits_0^\infty\!\!
e^{-\rho} L_{n-l-1}^{2l+1} (\rho)L_{n'-l'-1}^{2l'+1} (\rho)
\rho^{l+l'+3} d\rho.
\end{equation}
Using the recurrent relations and orthogonality
of the Legendre and Laguerre polynomials, we immediately find
\begin{equation}
J^0_{ll'}=\frac{l+1}{\sqrt{(2l+1)(2l+3)}}\, \delta _{l+1,l'}+
\frac{l}{\sqrt{(2l+1)(2l-1)}}\, \delta _{l-1,l'},
\label{N-el}
\end{equation}
\begin{eqnarray}
I_{nn'}^{l\,l+1}=-\sqrt{(n-l)(n+l+1)(n+l+2)(n+l+3)}\,\delta _{n+2,n'}
\nonumber\\
+2(2n-l)\sqrt{(n+l+1)(n+l+2)}\,\delta _{n+1,n'}
\nonumber\\
-6n\sqrt{n^2-(l+1)^2}\,\delta _{n,n'}
\nonumber\\
+2(2n+l)\sqrt{(n-l-1)(n-l-2)}\,\delta _{n-1,n'}
\nonumber\\
-\sqrt{(n+l)(n-l-1)(n-l-2)(n-l-3)}\,\delta _{n-2,n'},
\end{eqnarray}
and
\begin{equation}
I_{nn'}^{l\,l-1}=I_{n'n}^{l-1\,l}.
\end{equation}


\begin{references}
\bibitem{fn1} The solution of the unperturbed problem is denoted by
the upper script (0), see Sec.~\ref{sec-ptheory}.
\end{references}
\end{document}